# Partial Conway and iteration semirings

S.L. Bloom
Dept. of Computer Science
Stevens Institute of Technology
Hoboken, NJ, USA

Z. Ésik
Dept. of Computer Science
University of Szeged
Szeged, Hungary
GRLMC, Rovira i Virgili University
Tarragona, Spain

W. Kuich
Inst. for Discrete Mathematics and Geometry
Technical University of Vienna, Vienna, Austria

**Abstract**

A Conway semiring is a semiring $S$ equipped with a unary operation $^* : S \to S$, always called 'star', satisfying the sum star and product star identities. It is known that these identities imply a Kleene type theorem. Some computationally important semirings, such as $\mathbb{N}$ or $\mathbb{N}^{\mathrm{rat}}\langle\!\langle \Sigma^* \rangle\!\rangle$ of rational power series of words on $\Sigma$ with coefficients in $\mathbb{N}$, cannot have a total star operation satisfying the Conway identities. We introduce here *partial Conway semirings*, which are semirings $S$ which have a star operation defined only on an ideal of $S$; when the arguments are appropriate, the operation satisfies the above identities. We develop the general theory of partial Conway semirings and prove a Kleene theorem for this generalization.

## 1 Introduction

It is well-known that there exists no finite base of identities for the regular languages equipped with the operations of union $+$, product $\cdot$ and (Kleene) star $^*$; cf. [6, 18, 19]. The notion of Conway semirings involves two important identities for the star operation: the sum star and the product star identities,

$$\begin{aligned}(a+b)^* &= a^*(ba^*)^* \\ (ab)^* &= 1 + a(ba)^*b.\end{aligned}$$

It has been shown that Kleene's theorem for languages and automata, as well as its generalization to weighted automata, are consequences of these identities. Thus, it is possible to derive Kleene's theorem by purely equational reasoning from the axioms of Conway semirings, cf. [6, 3, 13]. Important examples of Conway semirings are

- the boolean semiring $\mathbb{B}$;
- the semirings $\mathbb{B}^{\mathrm{rat}}\langle\!\langle \Sigma^* \rangle\!\rangle$ of rational power series with coefficients in $\mathbb{B}$, which are isomorphic copies of the semirings of regular languages,
- the continuous or complete semirings [6, 8, 3].

However, many computationally important semirings do not have a totally defined star operation satisfying the Conway identities. Some examples of such semirings are the semirings $S^{\mathrm{rat}}\langle\!\langle \Sigma^* \rangle\!\rangle$ of rational power series over $\Sigma$ with coefficients in the semiring $S$, where $S$ is either the semiring $\mathbb{N}$ of natural numbers or a nontrivial ring (if $1^* = 1^* \cdot 1 + 1$, then $0 = 1$). The semiring $\mathbb{N}$ can be embedded into a Conway semiring, namely the semiring $\mathbb{N}_\infty$ obtained by adding a point $\infty$. By means of this embedding, Kleene's theorem for Conway semirings becomes indirectly applicable to weighted finite automata over $\mathbb{N}$. On the other hand, such an embedding does not exist for all semirings, so that Kleene's theorem for Conway semirings does not cover weighted finite automata over such semirings.

In this paper, we introduce *partial Conway semirings* as a generalization of Conway semirings. In a partial Conway semiring $S$, the domain $D(S)$ of the star operation is an *ideal* of the semiring; further, when restricted to this domain, the sum star and product star identities hold. We prove a Kleene theorem for partial Conway semirings, and thus obtain a single unified result which is directly applicable in all of the above situations. We also outline the general theory of partial Conway semirings which parallels with the theory of Conway semirings. This general theory provides the background for the Kleene theorem mentioned above.

Moreover, we also introduce *partial iteration semirings*, which are partial Conway semirings satisfying Conway's group identities, cf. [6, 16]. We define *partial iterative semirings* as star semirings in which certain linear equations have unique solutions. We prove that partial iterative semirings are partial iteration semirings. As an application of this result, we show that for any semiring $S$ and set $\Sigma$, the power series semiring $S\langle\!\langle \Sigma^* \rangle\!\rangle$ is a partial iterative semiring and thus a partial iteration semiring.

The results of this paper are used in [4], where the semirings $\mathbb{N}^{\mathrm{rat}}\langle\!\langle \Sigma^* \rangle\!\rangle$ are characterized as the free partial iteration semirings, and the semirings $\mathbb{N}_\infty^{\mathrm{rat}}\langle\!\langle \Sigma^* \rangle\!\rangle$ as the free algebras in a subvariety of iteration semirings satisfying three additional simple identities.

## 2  Semirings

A *semiring* [14] is an algebra $S = (S, +, \cdot, 0, 1)$ such that $(S, +, 0)$ is a commutative monoid, $(S, \cdot, 1)$ is a monoid, moreover $0$ is an absorbing element with respect to multiplication and product distributes over sum:

$$\begin{aligned} 0 \cdot a &= 0 \\ a \cdot 0 &= 0 \\ a(b + c) &= ab + ac \\ (b + c)a &= ba + ca \end{aligned}$$

for all $a, b, c \in S$. The operation $+$ is called *sum* or *addition*, and the operation $\cdot$ is called *product* or *multiplication*. A semiring $S$ is called *idempotent* if

$$a + a = a$$

for all $a \in S$. A morphism of semirings preserves the sum and product operations and the constants $0$ and $1$. Since semirings are defined by identities, the class of all semirings is a variety (see e.g., [15]) as is the class of all idempotent semirings.

An important example of a semiring is the semiring $\mathbb{N} = (\mathbb{N}, +, \cdot, 0, 1)$ of natural numbers equipped with the usual sum and product operations, and an important example of an idempotent semiring is the boolean semiring $\mathbb{B}$ whose underlying set is $\{0, 1\}$ and whose sum and product operations are the operations $\vee$ and $\wedge$, i.e., disjunction and conjunction. Actually $\mathbb{N}$ and $\mathbb{B}$ are respectively the initial semiring and the initial idempotent semiring.

We end this section by describing three constructions on semirings. For more information on semirings, the reader is referred to Golan's book [14].

## 2.1 Polynomial semirings and power series semirings

Suppose that $S$ is a semiring and $\Sigma$ is a set. Let $\Sigma^*$ denote the free monoid of all words over $\Sigma$ including the empty word $\epsilon$. A *formal power series*, or just *power series* over $S$ in the (noncommuting) letters in $\Sigma$ is a function $s : \Sigma^* \to S$. It is a common practice to represent a power series $s$ as a formal sum $\sum_{w \in \Sigma^*} (s,w)w$, where the *coefficient* $(s,w)$ is $ws$, the value of $s$ on the word $w$. The *support* of a series $s$ is the set $\mathrm{supp}(s) = \{w : (s,w) \neq 0\}$. When $\mathrm{supp}(s)$ is finite, $s$ is called a *polynomial*. We let $S\langle\!\langle \Sigma^* \rangle\!\rangle$ and $S\langle \Sigma^* \rangle$ respectively denote the collection of all power series and polynomials over $S$ in the letters $\Sigma$.

We define the sum $s + s'$ and product $s \cdot s'$ of two series $s, s' \in S\langle\!\langle \Sigma^* \rangle\!\rangle$ as follows. For all $w \in \Sigma^*$,

$$\begin{aligned} (s+s', w) &= (s,w) + (s',w) \\ (s \cdot s', w) &= \sum_{uu'=w} (s,u)(s',u'). \end{aligned}$$

We may identify any element $s \in S$ with the series, in fact polynomial that maps $\epsilon$ to $s$ and all other elements of $\Sigma^*$ to 0. In particular, 0 and 1 may be viewed as polynomials. It is well-known that equipped with the above operations and constants, $S\langle\!\langle \Sigma^* \rangle\!\rangle$ is a semiring which contains $S\langle \Sigma^* \rangle$ as a subsemiring.

The semiring $S\langle \Sigma^* \rangle$ can be characterized by a universal property. Consider the natural embedding of $\Sigma$ into $S\langle \Sigma^* \rangle$ such that each letter $\sigma \in \Sigma$ is mapped to the polynomial whose support is $\{\sigma\}$ which maps $\sigma$ to 1. By this embedding, we may view $\Sigma$ as a subset of $S\langle \Sigma^* \rangle$. Recall also that each $s \in S$ is identified with a polynomial. The following fact is well-known.

THEOREM 2.1 *Given any semiring $S'$, any semiring morphism $h_S : S \to S'$ and any function $h : \Sigma \to S'$ such that*

$$(sh_S)(ah) = (ah)(sh_S) \tag{1}$$

*for all $a \in \Sigma$ and $s \in S$, there is a unique semiring morphism $h^\sharp : S\langle \Sigma^* \rangle \to S'$ which extends both $h_S$ and $h$.*

The condition (1) means that for any $s \in S$ and letter $a \in \Sigma$, $sh_S$ commutes with $ah$. In particular, since $\mathbb{N}$ is initial, and since when $S = \mathbb{N}$ the condition (1) holds automatically, we obtain that any map $\Sigma \to S'$ into a semiring $S'$ extends to a unique semiring morphism $\mathbb{N}\langle \Sigma^* \rangle \to S'$, i.e., the polynomial semiring $\mathbb{N}\langle \Sigma^* \rangle$ is freely generated by $\Sigma$ in the variety of semirings. In the same way, $\mathbb{B}\langle \Sigma^* \rangle$ is freely generated by $\Sigma$ in the variety of idempotent semirings. Note that a series in $\mathbb{B}\langle\!\langle \Sigma^* \rangle\!\rangle$ may be identified with its support. Thus a series in $\mathbb{B}\langle\!\langle \Sigma^* \rangle\!\rangle$ corresponds to a language over $\Sigma$ and a polynomial in $\mathbb{B}\langle \Sigma^* \rangle$ to a finite language. The sum operation corresponds to set union and the product operation to concatenation. The constants 0 and 1 are the empty set and the singleton set $\{\epsilon\}$.

## 2.2 Matrix semirings and matrix theories

When $S$ is a semiring, then for each $n \geq 0$ the set $S^{n \times n}$ of all $n \times n$ matrices over $S$ is also a semiring. The sum operation is defined pointwise and product is the usual matrix product. The constants are the matrix $0_{nn}$ all of whose entries are 0 (often denoted just 0), and the diagonal matrix $E_n$ whose diagonal entries are all 1.

In addition to square matrices, we will also have opportunity to consider rectangular matrices of arbitrary size. A nice framework that arises with rectangular matrices is that of a matrix theory. Let $S$ be a semiring. The *matrix theory over $S$* [10, 3] is the category $\mathbf{Mat}_S$ whose objects are the natural numbers and whose morphisms $n \to p$ are the $n \times p$ matrices over $S$, i.e., the elements of the semiring $S^{n \times p}$. Composition is matrix product with the matrices $E_n$ being the identity morphisms. Equipped with the pointwise sum operation and the zero matrix $0_{np}$ all of whose entries are 0, each hom-set $S^{n \times p}$ of $\mathbf{Mat}_S$ is a commutative monoid. Moreover, composition distributes over finite sums. In the category $\mathbf{Mat}_S$, each object $n$ is both the categorical $n$-fold product of object 1 with itself and the $n$-fold coproduct of object 1 with itself. The canonical coproduct injections $1 \to n$ are the $1 \times n$ matrices $e_i$, $i = 1, \ldots, n$, having a 1 in the $i$th position and 0 elsewhere. The canonical projection morphisms $n \to 1$ are the transposes $e_i^T$ of these matrices. In any matrix theory $\mathbf{Mat}_S$, we associate a morphism $\hat{\rho} : m \to n$ with any function $\rho : \{1, \ldots, m\} \to \{1, \ldots, n\}$ by defining the $(i,j)$th entry of $\hat{\rho}$ to be 1 if $i\rho = j$ and 0 otherwise. We will call $\hat{\rho}$ a *functional matrix* and write just $\rho$ instead of $\hat{\rho}$. An *injective functional matrix* is a functional matrix corresponding to an injective function, and a *permutation matrix* is a functional matrix corresponding to permutation. Let $\mathbf{Mat}_S$ and $\mathbf{Mat}_{S'}$ be matrix theories. A morphism $\mathbf{Mat}_S \to \mathbf{Mat}_{S'}$ is a functor that preserves objects and the canonical coproduct injections and projections. It follows that any morphism $\mathbf{Mat}_S \to \mathbf{Mat}_{S'}$ preserves the additive structure, and determines and is determined by a semiring morphism $S \to S'$. Thus, the category of matrix theories is equivalent to the category of semirings. By this equivalence, we may identify each matrix theory morphism $h : \mathbf{Mat}_S \to \mathbf{Mat}_{S'}$ with its restriction to the $1 \times 1$ matrices which is a semiring morphism $S \to S'$. The image of a matrix $(A_{ij})_{ij}$ under $h$ is then given by $(A_{ij}h)_{ij}$. An isomorphism of matrix theories is a matrix theory morphism which is bijective on hom-sets. For the above facts and a more abstract treatment of matrix theories the reader is referred to [10].

## 2.3 Duality

The *dual* of a semiring $S = (S, +, \cdot, 0, 1)$ is the semiring $S^d = (S, +, \circ, 0, 1)$ which has the same sum operation and constants as $S$ and whose product operation is defined by $a \circ b = b \cdot a$, in the reverse order. Note that $(S^d)^d = S$, for all semirings $S$. A *dual morphism* $h : S_1 \to S_2$ between semirings $S_1 \to S_2$ is a morphism $S_1 \to S_2^d$, or equivalently, a morphism $S_1^d \to S_2$. A *dual isomorphism* is a bijective dual morphism. Note that for any semiring $S$, the identity function over $S$ is a dual isomorphism $S \to S^d$.

Suppose that $\mathbf{Mat}_{S_1}$ and $\mathbf{Mat}_{S_2}$ are matrix theories. A *dual matrix theory morphism* $h : \mathbf{Mat}_{S_1} \to \mathbf{Mat}_{S_2}$ maps morphisms $m \to n$ to morphisms $n \to m$, i.e., matrices in $S_1^{m \times n}$ to matrices in $S_2^{n \times m}$, such that $(AB)h = (Bh)(Ah)$ whenever $A$ and $B$ are matrices of appropriate size, moreover $E_n h = E_n$ for each $n$. It is required that canonical injections are mapped to canonical projections and vice versa, so that $e_i h = e_i^T$ and $e_i^T h = e_i$, for all $1 \times n$ matrices $e_i$. It follows that a dual matrix theory morphism preserves the zero matrices and the additive structure. A dual isomorphism of matrix theories is bijective on each hom-set.

The dual isomorphism $S \to S^d$ can be lifted to matrix theories.

PROPOSITION 2.2 *For any semiring $S$, the matrix theory $\mathbf{Mat}_{S^d}$ is dually isomorphic to $\mathbf{Mat}_S$, a dual isomorphism $\mathbf{Mat}_S \to \mathbf{Mat}_{S^d}$ maps each $A \in S^{n \times p}$ to its transpose $A^T \in S^{p \times n}$. In particular, denoting composition in $\mathbf{Mat}_{S^d}$ by $\circ$ we have:*

$$\begin{aligned}
(A+B)^T &= A^T + B^T \\
0_n^T &= 0^n \\
(AB)^T &= B^T \circ A^T \\
E_n^d &= E_n
\end{aligned}$$

*Proof.* It is clear that for each $n, p$ the assignment $A \mapsto A^T$ defines a bijection from the set of morphisms $n \to p$ in $\mathbf{Mat}_S$ to the set of morphisms $p \to n$ in $\mathbf{Mat}_{S^d}$. Moreover, the additive structure and the identities $E_n$ are preserved, and coproduct injections are mapped to coproduct projections and vice versa. Thus, it remains to prove that $(A \cdot B)^T = B^T \circ A^T$ holds for all $A \in S^{n \times p}$ and $B \in S^{p \times q}$ in $\mathbf{Mat}_S$, where composition (i.e., matrix product) in $\mathbf{Mat}_S$ is denoted $\cdot$ and matrix product in $\mathbf{Mat}_{S^d}$ is denoted $\circ$. But for all appropriate $i, j$, the $(i, j)$th entry of $(A \cdot B)^T$ is

$$(A \cdot B)^T_{ij} = (A \cdot B)_{ji} = \sum_k A_{jk} \cdot B_{ki} = \sum_k B_{ki} \circ A_{jk} = \sum_k B^T_{ik} \circ A^T_{kj} = (B^T \circ A^T)_{ij}.$$

□

REMARK 2.3 When $A$ or $B$ is a 0-1 matrix, or more generally, when each entry of $A$ commutes with any entry of $B$, then we have $(AB)^T = B^T A^T$.

## 3 Partial Conway semirings

The definition of Conway semirings involves two important identities of regular languages. Conway semirings appear implicitly in Conway [6] and were first defined explicitly in [2, 3]. See also [17]. On the other hand, the applicability of Conway semirings is limited due to the fact that the star operation is total, whereas many important semirings only have a partially defined star operation. Moreover, it is not true that all such semirings can be embedded into a Conway semiring with a totally defined star operation.

DEFINITION 3.1 *A partial $^*$-semiring is a semiring $S$ equipped with a partially defined star operation $^* : S \to S$ whose domain is an* ideal *of $S$. A $^*$-semiring is a partial $^*$-semiring $S$ such that $^*$ is defined on the whole semiring $S$. A morphism $S \to S'$ of (partial) $^*$-semirings is a semiring morphism $h : S \to S'$ such that for all $s \in S$, if $s^*$ is defined then so is $(sh)^*$ and $s^*h = (sh)^*$.*

Thus, in a partial $^*$-semiring $S$, $0^*$ is defined, and if $a^*$ and $b^*$ are defined then so is $(a+b)^*$, finally, if $a^*$ or $b^*$ is defined, then so is $(ab)^*$. When $S$ is a partial $^*$-semiring, we let $D(S)$ denote the domain of definition of the star operation.

DEFINITION 3.2 *A partial Conway semiring is a partial $^*$-semiring $S$ satisfying the following two axioms:*

1. *Sum star identity:*

$$(a+b)^* = a^*(ba^*)^* \qquad (2)$$

   *for all $a, b \in D(S)$.*

2. Product star identity:

$$(ab)^* = 1 + a(ba)^*b, \tag{3}$$

for all $a, b \in S$ such that $a \in D(S)$ or $b \in D(S)$.

A *Conway semiring* is a partial Conway semiring $S$ which is a *-semiring (i.e., $D(S) = S$). A morphisms of (partial) Conway semirings is a (partial) *-semiring morphism.

Note that in any partial Conway semiring $S$,

$$aa^* + 1 = a^* \tag{4}$$
$$a^*a + 1 = a^* \tag{5}$$
$$0^* = 1 \tag{6}$$

for all $a \in D(S)$. Moreover, if $a \in D(S)$ or $b \in D(S)$, then

$$(ab)^*a = a(ba)^*. \tag{7}$$

It follows that also

$$aa^* = a^*a \tag{8}$$
$$(a+b)^* = (a^*b)^*a^* \tag{9}$$

for all $a, b \in D(S)$. When $a \in D(S)$ we will denote $aa^* = a^*a$ by $a^+$ and call $^+$ the *plus* operation.

Conway semirings give rise to Conway matrix theories [3]. In the same way, partial Conway semirings give rise to partial Conway matrix theories defined below. We say that a collection $J$ of matrices in $\mathbf{Mat}_S$ is a *matrix ideal* if for any integers $m, n$, it contains the zero matrix $0_{mn}$ and is closed under sum, moreover, it is closed under multiplication with any matrix: if $A : m \to n$ in $J$ then for any $B : p \to m$ and $C : n \to q$ it holds that $BA, AC \in J$. It is easy to show that if $I$ is an ideal of $S$, then the collection of all matrices $J = M(I)$, all of whose entries are in $I$ is a matrix ideal of $\mathbf{Mat}_S$, and that any matrix ideal is of this sort. Thus, any matrix ideal of $\mathbf{Mat}_S$ is uniquely determined by an ideal of $S$.

DEFINITION 3.3 *Suppose that $S$ is a semiring and consider the matrix theory $\mathbf{Mat}_S$. We say that $\mathbf{Mat}_S$ is a* partial Conway matrix theory *if it is equipped with a star operation $A \mapsto A^*$, defined on the square matrices $A : n \to n$, $n \geq 0$ whose domain is the collection of all square matrices in a matrix ideal $M(I)$, moreover, the matrix versions of the sum and product star identities hold:*

$$(A + B)^* = A^*(BA^*)^* \tag{10}$$

*for all $A, B \in M(I)$, $A, B : n \to n$, and*

$$(AB)^* = E_n + A(BA)^*B, \tag{11}$$

*for all $A, B \in M(I)$, $A : n \to m$, $B : m \to n$. When $\mathbf{Mat}_S$ is a partial matrix theory such that star is defined on all square matrices, then we call $\mathbf{Mat}_S$ a* Conway matrix theory*. A morphism of (partial) matrix theories is a matrix theory morphism which preserves star.*

Note the following special cases of (11):

$$A^* = AA^* + E_n \tag{12}$$
$$A^* = A^*A + E_n \tag{13}$$
$$0^*_{nn} = E_n \tag{14}$$

where $A : n \to n$ in $M(I)$, $n \geq 0$. Also, $AA^* = A^*A$ for all $A : n \to n$ in $M(I)$. Below we will denote $AA^*$ by $A^+$.

If $\mathbf{Mat}_S$ is a (partial) Conway matrix theory, then by identifying a $1 \times 1$ matrix $(a)$ in $\mathbf{Mat}_S$ with the element $a$, the semiring $S$ becomes a (partial) Conway semiring. Conversely, any (partial) Conway semiring determines a (partial) Conway matrix theory, as we show below.

DEFINITION 3.4 *Suppose that $S$ is a partial Conway semiring with $D(S) = I$. We define a partial star operation on the semirings $S^{k \times k}$, $k \geq 0$, whose domain of definition is $I^{k \times k}$, the ideal of those $k \times k$ matrices all of whose entries are in $I$. When $k = 0$, $S^{k \times k}$ is trivial as is the definition of star. When $k = 1$, we use the star operation on $S$. Assuming that $k > 1$ we write $k = n + 1$. For a matrix $\begin{pmatrix} a & b \\ c & d \end{pmatrix}$ in $I^{k \times k}$, define*

$$\begin{pmatrix} a & b \\ c & d \end{pmatrix}^* = \begin{pmatrix} \alpha & \beta \\ \gamma & \delta \end{pmatrix} \quad (15)$$

*where $a \in S^{n \times n}$, $b \in S^{n \times 1}$, $c \in S^{1 \times n}$ and $d \in S^{1 \times 1}$, and where*

$$\begin{aligned} \alpha &= (a + bd^*c)^* & \beta &= \alpha b d^* \\ \gamma &= \delta c a^* & \delta &= (d + ca^*b)^*. \end{aligned}$$

By the above definition, we have also defined a star operation on those square matrices in $\mathbf{Mat}_S$ which belong to $M(I)$. It is known (cf. [3]) that when $S$ is a Conway semiring, then equipped with the above star operation, $\mathbf{Mat}_S$ is a Conway matrix theory. More generally, but with the same proof, we have:

THEOREM 3.5 *Suppose that $S$ is a partial Conway semiring with $D(S) = I$. Then, equipped with the above star operation, $\mathbf{Mat}_S$ is a partial Conway matrix theory where the star operation is defined on the square matrices in $M(I)$.*

COROLLARY 3.6 *If $S$ is a (partial) Conway semiring, then so is the semiring $S^{n \times n}$, for each $n$.*

COROLLARY 3.7 *The category of (partial) Conway semirings is equivalent to the category of (partial) Conway matrix theories.*

Also the following result is known to hold for Conway matrix theories.

THEOREM 3.8 *Suppose that $\mathbf{Mat}_S$ is a partial Conway matrix theory where star is defined on the square matrices in $M(I)$. Then the following identities hold.*

1. *The* matrix star identity *(15) for all possible decompositions of a square matrix in $M(I)$ into four blocks such that $a$ and $d$ are square matrices, i.e., where $a : n \to n$, $b : n \to m$, $c : m \to n$ and $d : m \to m$.*

2. *The* permutation identity *(16)*

$$(\pi A \pi^T)^* = \pi A^* \pi^T, \quad (16)$$

*for all $A : n \to n$ in $M(I)$ and any permutation matrix $\pi : n \to n$, where $\pi^T$ denotes the transpose of $\pi$.*

The proof is the same as for Conway matrix theories, cf. [3]. For later use we note the following. When $\mathbf{Mat}_S$ is a partial Conway matrix theory with star operation defined on the square matrices in $M(I)$, and if $A = \begin{pmatrix} a & b \\ c & d \end{pmatrix}$ is a matrix with entries in $M(I)$, partitioned as above, then

$$A^+ = \begin{pmatrix} (a+bd^*c)^+ & (a+bd^*c)^*bd^* \\ (d+ca^*b)^*ca^* & (d+ca^*b)^+ \end{pmatrix} \tag{17}$$

$$A^* = \begin{pmatrix} (a+bd^*c)^* & a^*b(d+ca^*b)^* \\ d^*c(a+bd^*c)^* & (d+ca^*b)^* \end{pmatrix} \tag{18}$$

## 3.1 Duality

Suppose that $S$ is a partial $*$-semiring. Then we may equip $S^d$ with the same star operation. Since $D(S)$ is also an ideal of $S^d$, we have that $S^d$ is a partial $*$-semiring. Let $S$ and $S'$ be $*$-semirings. We say that a function $h : S \to S'$ is a dual morphism of partial $*$-semirings if it is dual semiring morphism mapping $D(S)$ to $D(S')$ which preserves star. A dual isomorphism is a bijective dual morphism.

PROPOSITION 3.9 *When $S$ is a partial Conway semiring, so is $S^d$. Moreover, the identity function $S \to S$ is a dual isomorphism $S \to S^d$.*

*Proof.* This follows from the fact that (9) holds in all partial Conway semirings. □

Since for partial Conway semirings $S$, the semiring $S^d$ is also a partial Conway semiring, $\mathbf{Mat}_{S^d}$ is a partial Conway matrix theory. A dual morphism $\mathbf{Mat}_S \to \mathbf{Mat}'_S$ between partial Conway matrix theories also preserves star. A dual isomorphism is a dual morphism which is bijective on hom-sets.

PROPOSITION 3.10 *Suppose that $\mathbf{Mat}_S$ is a partial Conway matrix theory. Then the function $A \mapsto A^T$, $A : m \to n$ in $\mathbf{Mat}_S$, is a dual isomorphism $\mathbf{Mat}_S \to \mathbf{Mat}_{S^d}$ of partial Conway matrix theories.*

*Proof.* We know that the assignment $A \mapsto A^T$ defines a dual isomorphism of the underlying matrix theories. Let $I = D(S) = D(S^d)$. It is clear that if $A$ is in $M(I)$ then $A^T$ is also in $M(I)$. To complete the proof, we still have to show that $(A^T)^\otimes = (A^*)^T$ for all square matrices in $M(I)$, where $^\otimes$ denotes the star operation in $\mathbf{Mat}_{S^d}$. To prove this, let $A : n \to n$ in $\mathbf{Mat}_S$. When $n = 0$ or $n = 1$, our claim is clear. We proceed by induction on $n$. Assume that $n > 1$. Then let us write $A = \begin{pmatrix} a & b \\ c & d \end{pmatrix}$, where $a$ and $d$ are square matrices of size $(n-1) \times (n-1)$ and $1 \times 1$, respectively. Then, using (15), (18) and the induction hypothesis,

$$\begin{aligned}
(A^T)^\circledast &= \begin{pmatrix} a^T & c^T \\ b^T & d^T \end{pmatrix}^\circledast \\
&= \begin{pmatrix} (a^T + c^T \circ (d^T)^\circledast \circ b^T)^\circledast & (a^T + c^T \circ (d^T)^\circledast \circ b^T)^\circledast \circ c^T \circ (d^T)^\circledast \\ (d^T + b^T \circ (a^T)^\circledast \circ c^T)^\circledast \circ b^T \circ (a^T)^\circledast & (d^T + b^T \circ (a^T)^\circledast \circ c^T)^\circledast \end{pmatrix} \\
&= \begin{pmatrix} ((a+bd^*c)^*)^T & (d^*c(a+bd^*c)^*)^T \\ (a^*b(d+ca^*b)^*)^T & ((d+ca^*b)^*)^T \end{pmatrix} \\
&= \begin{pmatrix} (a+bd^*c)^* & a^*b(d+ca^*b)^* \\ d^*c(a+bd^*c)^* & (d+ca^*b)^* \end{pmatrix}^T \\
&= \left( \begin{pmatrix} a & c \\ b & d \end{pmatrix}^* \right)^T \\
&= (A^*)^T. \quad \square
\end{aligned}$$

## 4 Partial iteration semirings

Many important (partial) Conway semirings satisfy the group identities associated with the finite groups, introduced by Conway [6]. Such *-semirings are the continuous *-semirings, or more generally, the inductive *-semirings of [12], the *-semirings that arise from complete semirings [3], or the (partial) iterative semirings defined in the next section. When a (partial) Conway semiring satisfies the group identities, it will be called a (partial) iteration semiring.

DEFINITION 4.1 *We say that the* group identity associated with a finite group $G$ of order $n$ holds *in a partial Conway semiring $S$ if*

$$e_1 M_G^* u_n = (a_1 + \cdots + a_n)^* \tag{19}$$

*holds, where* $a_1, \cdots, a_n$ *are arbitrary elements in $D(S)$, and where $M_G$ is the $n \times n$ matrix whose $(i,j)$th entry is $a_{i^{-1}j}$, for all $1 \le i, j \le n$, and $e_1$ is the $1 \times n$ 0-1 matrix whose first entry is 1 and whose other entries are 0, finally $u_n$ is the $n \times 1$ matrix all of whose entries are 1.*

Identity (19) asserts that the sum of the entries of the first row of $M_G^*$ is $(a_1 + \cdots + a_n)^*$. For example, the group identity associated with the group of order 2 is

$$\begin{pmatrix} 1 & 0 \end{pmatrix} \begin{pmatrix} a_1 & a_2 \\ a_2 & a_1 \end{pmatrix}^* \begin{pmatrix} 1 \\ 1 \end{pmatrix} = (a_1 + a_2)^*$$

which by the matrix star identity can be written as

$$(a_1 + a_2 a_1^* a_2)^*(1 + a_2 a_1^*) = (a_1 + a_2)^*.$$

(It is known that in Conway semirings, this identity is further equivalent to $(a^2)^*(1+a) = a^*$.)

DEFINITION 4.2 *We say that a Conway semiring $S$ is an* iteration semiring *if it satisfies all group identities. We say that a partial Conway semiring $S$ is a* partial iteration semiring *if it satisfies all group identities (19) where $a_1, \cdots, a_n$ range over $D(S)$. A morphism of (partial) iteration semirings is a (partial) Conway semiring morphism.*

*We say that a (partial) Conway matrix theory is a* (partial) matrix iteration theory *if it satisfies all group identities. A morphism of (partial) matrix iteration theories is a (partial) Conway matrix theory morphism.*

It is clear that a (partial) Conway semiring $S$ is a (partial) iteration semiring iff $\mathbf{Mat}_S$ is a (partial) iteration semiring. Also, the category of (partial) iteration semirings is equivalent to the category of (partial) matrix iteration theories.

PROPOSITION 4.3 *Suppose that the partial Conway semiring $S$ satisfies the group identity (19) for all $a_1, \cdots, a_n \in D(S)$. Then $S$ also satisfies*

$$u_n^T M_G^* e_1^T = (a_1 + \cdots + a_n)^*, \qquad (20)$$

*for all $a_1, \cdots, a_n \in D(S)$, where $e_1$, $M_G$ and $u_n$ are defined as above. Thus, if $S$ is an iteration semiring, then (20) holds for all finite groups $G$.*

Proof. For each $i \in \{1, \ldots, n\}$, let $\pi_i$ denote the permutation matrix corresponding to the bijection $\{1, \cdots, n\} \to \{1, \cdots, n\}$, $j \mapsto ij$, where the product $ij$ is computed in the group $G$. An easy calculation shows that $\pi M_G \pi^T = M_G$. Thus, by the permutation identity, also $\pi M_G^* \pi^T = M_G^*$. Since this holds for all $i$, also $(M_G^*)_{i1} = (M_G^*)_{1i^{-1}}$ for all $i$. Thus, the entries of the first column of $M_G^*$ form a permutation of the entries of the first row of $M_G^*$. We conclude that if (19) holds, then so does (20). $\square$

REMARK 4.4 In Conway semirings, the group identity (19) is equivalent to (20).

The group identities seem to be extremely difficult to verify in practice. However, they are implied by the simpler functorial star conditions defined below.

DEFINITION 4.5 *Suppose that $S$ is a partial Conway semiring so that $\mathbf{Mat}_S$ is a partial Conway matrix theory. Let $I = D(S)$, and let $\mathcal{C}$ be a class of matrices in $\mathbf{Mat}_S$. We say that $\mathbf{Mat}_S$ has a* functorial star *with respect to $\mathcal{C}$ if for all $A : m \to m$ and $B : n \to n$ in $M(I)$ and for all $C : m \to n$ in $\mathcal{C}$, if $AC = CB$ then $A^*C = CB^*$.*

Suppose that $\mathcal{C}$ is a class of matrices in a Conway matrix theory $\mathbf{Mat}_S$. Then let $B(\mathcal{C})$ denote the class of block diagonal rectangular matrices whose diagonal blocks are in $\mathcal{C}$.

LEMMA 4.6 *If a Conway matrix theory $\mathbf{Mat}_S$ has a functorial star with respect to $\mathcal{C}$, then it also has a functorial star with respect to the class $B(\mathcal{C})$.*

Proof. It suffices to prove the following. Let $A : m \to m$ and $B : n \to n$ and $C : m \to n$ in $\mathbf{Mat}_S$ with $AC = CB$ such that $A^*$ and $B^*$ are defined. Moreover, suppose that $C = \begin{pmatrix} c_1 & 0 \\ 0 & c_2 \end{pmatrix}$ where $c : m_1 \to n_1$, $d : m_2 \to n_2$ with $m_1 + m_2 = m$ and $n_1 + n_2 = n$. If $\mathbf{Mat}_S$ has a functorial star with respect to $\{c, d\}$, then $\mathbf{Mat}_S$ has a functorial star with respect to $\{C\}$. To prove this, let us write

$$A = \begin{pmatrix} a_1 & a_2 \\ a_3 & a_4 \end{pmatrix}, \quad B = \begin{pmatrix} b_1 & b_2 \\ b_3 & b_4 \end{pmatrix}$$

where $a_1 : m_1 \to m_1$, etc. Since $AC = CB$, we have

$$\begin{array}{ll} a_1 c = c b_1 & a_2 d = c b_2 \\ a_3 c = d b_3 & a_4 d = d b_4 \end{array}$$

Thus, since $\mathbf{Mat}_S$ has a functorial star with respect to $\{c, d\}$,

$$a_1^* c = c b_1^* \quad \text{and} \quad a_4^* d = d b_4^*.$$

Using these equations, it follows that

$$(a_1 + a_2 a_3^* a_4)c = c(b_1 + b_2 b_3^* b_4)$$
$$(a_4 + a_3 a_1^* a_2)d = d(b_4 + b_3 b_1^* b_2)$$

Thus, using again the fact that $\mathbf{Mat}_S$ has a functorial star with respect to $\{c, d\}$, it follows that

$$(a_1 + a_2 a_3^* a_4)^* c = c(b_1 + b_2 b_3^* b_4)^*$$
$$(a_1 + a_2 a_3^* a_4)^* a_2 a_4^* d = c(b_1 + b_2 b_3^* b_4)^* b_2 b_4^*$$
$$(a_4 + a_3 a_1^* a_2)^* d = d(b_4 + b_3 b_1^* b_2)^*$$
$$(a_4 + a_3 a_1^* a_2)^* a_3 a_1^* c = d(b_4 + b_3 b_1^* b_2)^* b_3 b_1^*,$$

so that $A^* C = C B^*$. □

PROPOSITION 4.7 *Suppose that $\mathbf{Mat}_S$ is a (partial) Conway matrix theory.*

1. *$\mathbf{Mat}_S$ has a functorial star with respect to the class of all injective functional matrices and their transposes.*

2. *If $\mathbf{Mat}_S$ has a functorial star with respect to the class of functional matrices $m \to 1$, $m \geq 2$, then $\mathbf{Mat}_S$ has a functorial star with respect to the class of all functional matrices.*

3. *If $\mathbf{Mat}_S$ has a functorial star with respect to the class of transposes of functional matrices $m \to 1$, $m \geq 2$, then $\mathbf{Mat}_S$ has a functorial star with respect to the class of transposes of all functional matrices.*

4. *If $\mathbf{Mat}_S$ has a functorial star with respect to the class of all functional matrices $m \to 1$, $m \geq 2$, then $\mathbf{Mat}_S$ is a partial matrix iteration theory.*

*Proof.* The fact that when $\mathbf{Mat}_S$ is a Conway matrix theory, then $\mathbf{Mat}_S$ has a functorial dagger with respect to the class of injective functional matrices and their transposes is proved in [3]. The same proof applies for partial Conway matrix theories. The second and third claims follow from the preceding lemma. The last fact is proved as follows. Let $G$ be a finite group of order $n$ and consider the matrix $M_G$ defined above. Using the notation in (19), we have $M_G u_n = u_n a$ where $a$ denotes the sum $a_1 + \cdots + a_n$. Thus, if $\mathbf{Mat}_S$ has a functorial star with respect to all functional matrices $m \to 1$, $m \geq 2$, then $M_G^* u_n = u_n a^*$ and $e_1 M_G^* u_n = a^*$. Since also $a u_n^T = u_n^T M_G$, if $\mathbf{Mat}_S$ has a functorial star with respect to all transposes of functional matrices $m \to 1$, $m \geq 2$, then $a^* u_n^T = u_n^T M_G^*$ and $a^* = u_n^T M_G^* e_1^T$. But this implies that $e_1 M_G^* u_n = a^*$. (See Remark 4.4.)
□

An important identity that holds in all iteration semirings $S$ and matrix iteration theories $\mathbf{Mat}_S$ is the *commutative identity*, cf. [3], which is a generalization of the group identities. It allows us to deduce $A^* \rho = \rho B^*$ from $A \rho = \rho B$ under certain conditions, where $A$ and $B$ are square matrices and $\rho$ is a functional matrix. The commutative identity also holds in partial matrix iteration theories (with the obvious restriction on the applicability of the star operation). Since the dual of an iteration semiring is also an iteration semiring, see below, the *dual commutative identity* of [3] also holds in (partial) matrix iteration theories.

## 4.1 Duality

PROPOSITION 4.8 *Suppose that $S$ is a partial Conway semiring. Then $S$ is a partial iteration semiring iff $S^d$ is. Thus, $\mathbf{Mat}_S$ is a partial matrix iteration theory iff $\mathbf{Mat}_{S^d}$ is.*

*Proof.* Suppose that $\mathbf{Mat}_S$ is a partial Conway matrix theory with star operation defined on the square matrices in the matrix ideal $M(I)$. Let $M_G = M_G(a_1, \ldots, a_n)$ be the matrix associated with the finite group $G$ of order $n$, see Definition 4.1, where $a_1, \cdots, a_n$ are in $I$. Note that $M_G^T$ is just $M_G(a_{1^{-1}}, \cdots, a_{n^{-1}})$, the matrix obtained from $M_G$ by replacing each occurrence of $a_i$ with $a_{i^{-1}}$. Since $\mathbf{Mat}_S$ is a matrix iteration theory, the group identity associated with $G$ holds in $\mathbf{Mat}_S$. In particular, $e_1 (M_G^T)^* u_n = (a_{1^{-1}} + \cdots + a_{n^{-1}})^* = (a_1 + \cdots + a_n)^*$ holds. Thus, by Proposition 4.3,

$$\begin{aligned} e_1 \circ M_G^\otimes \circ u_n &= (u_n^T (M_G^\otimes)^T e_1^T)^T \\ &= (u_n^T (M_G^T)^* e_1^T)^T \\ &= u_n^T (M_G^T)^* e_1^T \\ &= (a_1 + \cdots + a_n)^*. \quad \square \end{aligned}$$

## 5 Partial iterative semirings

In this section we exhibit a class of partial iteration semirings.

DEFINITION 5.1 *A partial iterative semiring is a partial $^*$-semiring $S$ such that for every $a \in D(S)$ and $b \in S$, $a^*b$ is the unique solution of the equation $x = ax + b$. A morphism of partial iterative semirings is a $^*$-semiring morphism.*

We note that any semiring $S$ with a distinguished ideal $I$ such that for all $a \in I$ and $b \in S$, the equation $x = ax + b$ has a unique solution can be turned into a partial iterative semiring, where star is defined on $I$. Indeed, when $a \in I$, define $a^*$ as the unique solution of the equation $x = ax + 1$. It follows that $aa^*b + b = a^*b$ for all $b$, so that $a^*b$ is the unique solution of $x = ax + b$. We also note that when $S, S'$ are partial iterative semirings, then any semiring morphism $h : S \to S'$ with $D(S)h \subseteq D(S')$ automatically preserves star. Indeed, when $a \in D(S)$, then $a^* = aa^* + 1$, thus $a^*h = (ah)(a^*h) + 1$, showing that $a^*h$ is a solution of the equation $x = (ah)x + 1$ over $S'$. But since $ah$ is in $D(S')$, the only solution is $(ah)^*$. Thus, $a^*h = (ah)^*$.

PROPOSITION 5.2 *Every partial iterative semiring is a partial Conway semiring.*

*Proof.* Suppose that $S$ is a partial iterative semiring and $a, b \in D(S)$. Since $a + b \in D(S)$ and

$$\begin{aligned} (a+b)a^*(ba^*)^* + 1 &= aa^*(ba^*)^* + ba^*(ba^*)^* + 1 \\ &= aa^*(ba^*)^* + (ba^*)^* \\ &= (aa^* + 1)(ba^*)^* \\ &= a^*(ba^*)^*, \end{aligned}$$

it follows by uniqueness of solutions that $(a+b)^* = a^*(ba^*)^*$. Also, if $a$ or $b$ is in $D(S)$, then $ab \in D(S)$ and

$$\begin{aligned} ab(a(ba)^*b + 1) + 1 &= a(ba(ba)^* + 1)b + 1 \\ &= a(ba)^*b + 1, \end{aligned}$$

so that $(ab)^* = a(ba)^*b + 1$, by uniqueness. $\square$

COROLLARY 5.3 *If $S$ is a partial iterative semiring, then $\mathbf{Mat}_S$, equipped with the star operation defined on matrices in Section 3, is a partial Conway matrix theory.*

It is known, cf. [5, 3], that if a class of functions in several variables over a set has certain closure properties, and if each fixed point equation with respect to a function in the class has a unique solution, then the same holds for finite systems of fixed point equations involving functions from the class. Moreover, such systems can be solved by successive elimination of the unknowns. A specialization of this result is given below.

THEOREM 5.4 *Suppose that $S$ is a partial iterative semiring with $D(S) = I$. Then the following holds in the partial Conway matrix theory $\mathbf{Mat}_S$. For any $A : n \to n$ in $M(I)$ and any $B : n \to p$, $A^*B$ is the unique solution of the matrix equation $X = AX + B$.*

*Proof.* We provide a proof for completeness. Let $A$ and $B$ be matrices as above. Since $\mathbf{Mat}_S$ is a partial Conway matrix theory, $AA^*B + B = (AA^* + E_n)B = A^*B$ by (12). To complete the proof, we have to show that the solution is unique. This is clear when $n = 0$ or $n = 1$. We proceed by induction on $n$. Assuming $n > 1$, write $A$ in the form

$$A = \begin{pmatrix} a & b \\ c & d \end{pmatrix}$$

where $a$ and $d$ are square matrices of size $m \times m$ and $k \times k$, respectively, where $m, k > 0$, $m + k = n$. Then let

$$X = \begin{pmatrix} x \\ y \end{pmatrix} \quad B = \begin{pmatrix} e \\ f \end{pmatrix}$$

where $x, e$ are of size $m \times p$ and $y, f$ are of size $k \times p$. Using he above decomposition of the matrices, we can write the equation $X = AX + B$ as

$$x = ax + by + e \tag{21}$$
$$y = cx + dy + f. \tag{22}$$

By uniqueness, $x = a^*(by + e)$ and $y = d^*(cx + f)$. Substituting the expression for $x$ in (21) and the expression for $y$ in (22) gives

$$x = (a + bd^*c)x + e + bd^*f$$
$$y = (d + ca^*b)y + f + ca^*e.$$

But since $a + bd^*c$ and $d + ca^*b$ are in $M(I)$, each of these equations has a unique solution. □

The next fact follows from Theorem 5.4 and a result from [9]. For completeness, we provide a proof.

THEOREM 5.5 *Suppose that $S$ is a partial iterative semiring with $D(S) = I$ and $A : n \to n$ and $B : n \to p$ in $\mathbf{Mat}_S$ such that $A^k \in M(I)$ for some $k \geq 1$. Then the equation $X = AX + B$ has a unique solution $(A^k)^*(A^{k-1}B + \cdots + B)$.*

*Proof.* Let $f(X) = AX + B$. Then $f^m(X) = A^m X + A^{m-1}B + \cdots + B$ for all $m \geq 1$. By assumption, the equation $X = f^k(X)$ has a unique solution $X_0 = (A^k)^*(A^{k-1}B + \cdots + B)$. Our aim is to show that $X_0$ is the unique solution of $X = f(X)$. But $f(X_0) = f(f^k(X_0)) = f^k(f(X_0))$, and since $X_0$ is the unique solution of the equation $X = f^k(X)$, we conclude that $X_0 = f(X_0)$. Also, if $X = f(X)$, then $X = f^m(X)$ for all $m$, so that any solution of the equation $X = f(X)$ is a solution of the equation $X = f^k(X)$. □

REMARK 5.6 Suppose that $S$ is a partial iterative semiring with $D(S) = I$. Note that if $a \in S$ and $k \geq 1$ are such that $a^k \in I$, then $a^m \in I$ for all $m \geq k$. Let $J = \{a \in S : \exists k \geq 1 \ a^k \in I\}$, so that $I \subseteq J$. By Theorem 5.5, the equation $x = ax + b$ has a unique solution for each $a \in I$ and $b \in S$, and this unique solution can be written as $(a^k)^*(a^{k-1}b + \cdots + b)$ whenever $a^k \in I$. Now suppose that $S$ is commutative. Then $J$ is also an ideal of $S$. This follows by noting that $0 \in J$, moreover, if $a^k \in I$ and $b^k \in I$, then $(a+b)^{2k} \in I$. Moreover, if $a^k \in I$, then for any $b \in S$, $(ab)^k = a^k b^k \in I$. Thus, if we define $a^*$ for $a \in J$ as the unique solution of the equation $x = ax + 1$, then $S$ is a partial iterative semiring, where the domain of definition of the star operation is $J$. Moreover, this star operation agrees with the the original one on the ideal $I$.

Our next aim is to show that partial iterative semirings are partial iteration semirings and thus the matrix theories of partial iterative semirings are partial iteration matrix theories.

THEOREM 5.7 *Suppose that $S$ is a partial iterative semiring. Then the partial Conway matrix theory $\mathbf{Mat}_S$ has a functorial star with respect to all matrices. Thus, if $AC = CB$ for some matrices $A : n \to n$, $B : m \to m$ and $C : n \to m$, where $A, B \in M(I)$, then $A^*C = CB^*$.*

*Proof.* We have $ACB^* + C = CBB^* + C = CB^*$, showing that $CB^*$ is a solution of the equation $X = AX + C$. But the unique solution is $A^*C$. Thus $A^*C = CB^*$. □

COROLLARY 5.8 *Any partial iterative semiring is a partial iteration semiring.*

*Proof.* Let $S$ be a partial iterative semiring. We already know that $S$ is a partial Conway semiring (cf. Proposition 5.2) and thus $\mathbf{Mat}_S$ is a partial Conway matrix theory. By Theorem 5.7, $\mathbf{Mat}_S$ has a functorial star with respect to all matrices. Thus, by Proposition 4.7, $\mathbf{Mat}_S$ is a partial matrix iteration theory. □

We give an application of the above corollary. Let $S$ be a semiring and $\Sigma$ a set, and consider the power series semiring $S\langle\!\langle \Sigma^* \rangle\!\rangle$. Following [1], we call a series $s \in S\langle\!\langle \Sigma^* \rangle\!\rangle$ *proper* if $(s, \epsilon) = 0$. Clearly, the proper series form an ideal. It is known, cf. [1], that for any series $s, r$, if $s$ is proper, then the equation $x = sx + r$ has a unique solution. Moreover, this unique solution is $s^*r$, where $s^*$ is the unique solution of the equation $y = sy + 1$.

COROLLARY 5.9 *For any semiring $S$ and set $\Sigma$, $S\langle\!\langle \Sigma^* \rangle\!\rangle$, equipped with the above star operation defined on the proper series, is a partial iterative semiring and thus a partial iteration semiring.*

REMARK 5.10 Consider the above partial iterative semiring $S\langle\!\langle \Sigma^* \rangle\!\rangle$ with star operation defined on the ideal $I$ of proper series. Let $J$ be defined as in Remark 5.6. Then $J$ is the collection of all *cycle free* series, cf. [7]. As shown in Remark 5.6, if $S$ is commutative then $J$ is also an ideal, and the star operation can be extended to the ideal $J$ so that $S\langle\!\langle \Sigma^* \rangle\!\rangle$ becomes a partial iterative semiring with star defined on $J$. By the above Corollary, this partial iterative semiring is a partial iteration semiring.

REMARK 5.11 Suppose that $S$ is partial iterative semiring with star operation defined on $D(S) = I$, and suppose that $S_0$ is a subsemiring of $S$ which is equipped with a unary operation $^\otimes$. Moreover, suppose that $S$ is the direct sum of $S_0$ and $I$, so that each $s \in S$ has a unique representation as a sum $x + a$ with $x \in S_0$ and $a \in A$. It is shown in [3, 2] that if $S_0$, equipped with the operation $^\otimes$, is a Conway semiring, then there is a unique way to turn $S$ into a Conway semiring whose star operation extends $^\otimes$. This operation also extends the star operation originally defined on $I$. Moreover, when $S_0$ is an iteration semiring, then $S$ is also an iteration semiring. In particular, if $S$ is a Conway or an iteration semiring, then so is $S\langle\!\langle \Sigma^* \rangle\!\rangle$.

We end this section by defining iterative semirings.

DEFINITION 5.12 *An* iterative semiring *is a partial iterative semiring $S$ such that $D(S)$ is the collection of all elements $s \in S$ which cannot be written in the form $1 + s'$. A morphism of iterative semirings is a partial $^*$-semiring morphism.*

It follows from our results that every iterative semiring is a partial iteration semiring. For example, $\mathbb{N}\langle\!\langle \Sigma^* \rangle\!\rangle$ is an iterative semiring.

Question: Is there a (partial) iterative semiring whose dual is not iterative?

# 6 Kleene theorem

The classical Kleene theorem equates languages recognizable by finite automata with the regular languages, and its generalization by Schützenberger equates power series recognizable by finite weighted automata with rational power series. In this section we establish a Kleene theorem for partial Conway semirings. To this end, we define a general notion of (finite) automaton in partial Conway semirings.

DEFINITION 6.1 *Suppose that $S$ is a partial Conway semiring, $S_0$ is a subsemiring of $S$ and $\Sigma$ is a subset of $D(S)$. An* automaton *in $S$ over $(S_0, \Sigma)$ is a triplet $\mathbf{A} = (\alpha, A, \beta)$ consisting of an* initial vector $\alpha \in S_0^{1 \times n}$, *a* transition matrix $A \in S_0\Sigma^{n \times n}$, *where $S_0\Sigma$ is the set of all linear combinations over $\Sigma$ with coefficients in $S_0$, and a* final vector $\beta \in S_0^{n \times 1}$. *The integer $n$ is called the* dimension *of $\mathbf{A}$. The* behavior *of $\mathbf{A}$ is $|\mathbf{A}| = \alpha A^* \beta$.*

(Since $A \in D(S)^{n \times n}$, $A^*$ exists.)

DEFINITION 6.2 *We say that $s \in S$ is* recognizable *over $(S_0, \Sigma)$ is $s$ is the behavior of some automaton over $(S_0, \Sigma)$. We let $\mathbf{Rec}_S(S_0, \Sigma)$ denote the set of all elements of $S$ which are recognizable over $(S_0, \Sigma)$.*

Next we define rational elements.

DEFINITION 6.3 *Let $S, S_0$ and $\Sigma$ be as above. We say that $s \in S$ is* rational *over $(S_0, \Sigma)$ if $s = x + a$ for some $x \in S_0$ and some $a \in S$ which is contained in the least set $\mathbf{Rat}'_S(S_0, \Sigma)$ containing $\Sigma \cup \{0\}$ and closed under the* rational operations $+$, $\cdot$, $^+$ *and left and right multiplication with elements of $S_0$. We let $\mathbf{Rat}_S(S_0, \Sigma)$ denote the set of rational elements over $(S_0, \Sigma)$.*

Note that $\mathbf{Rat}'_S(S_0, \Sigma) \subseteq D(S)$ and that the element $a$ in the above definition is in $D(S)$.

PROPOSITION 6.4 *Suppose that $S$ is a partial Conway semiring, $S_0$ is a subsemiring of $S$ and $\Sigma$ is a subset of $D(S)$. Then $\mathbf{Rat}_S(S_0, \Sigma)$ contains $S_0$ and is closed under sum and product. Moreover, it is closed under star iff it is closed under the plus operation.*

*Proof.* Since $0 \in \mathbf{Rat}'_S(S_0, \Sigma)$, it follows that $S_0 \subseteq \mathbf{Rat}_S(S_0, \Sigma)$. Let $r = x + a$ and $s = y + b$ be in $\mathbf{Rat}_S(S_0, \Sigma)$, where $x, y \in S_0$ and $a, b \in \mathbf{Rat}'_S(S_0, \Sigma)$. Then $r + s = (x + y) + (a + b)$ and $rs = xy + (xb + ay + ab)$, so that $r + s$ and $rs$ are in $\mathbf{Rat}_S(S_0, \Sigma)$. Since $\mathbf{Rat}_S(S_0, \Sigma)$ is closed under sum and product and contains 1, it is closed under star iff it is closed under plus. □

The following fact is clear.

PROPOSITION 6.5 *Suppose that $S$ is a partial Conway semiring, $S_0$ is a subsemiring of $S$ and $\Sigma$ is a subset of $D(S)$. Then $\mathbf{Rat}_S(S_0, \Sigma)$ is contained in the least subsemiring of $S$ containing $S_0$ and $\Sigma$ which is closed under star.*

We give two sufficient conditions under which $\mathbf{Rat}_S(S_0, \Sigma)$ is closed under star.

PROPOSITION 6.6 *Let $S, S_0$ and $\Sigma$ be as above. Assume that either $S_0 \subseteq D(S)$ and $S_0$ is closed under star, or the following condition holds:*

$$\forall x \in S_0 \forall a \in D(S) \quad (x + a \in D(S) \Rightarrow x = 0). \tag{23}$$

*Then $\mathbf{Rat}_S(S_0, \Sigma)$ is closed under star. Moreover, in either case, $\mathbf{Rat}_S(S_0, \Sigma)$ is the least subsemiring of $S$ containing $S_0$ and $\Sigma$ which is closed under star.*

*Proof.* We know that $\mathbf{Rat}_S(S_0, \Sigma)$ is closed under star iff it is closed under plus.

Assume first that $S_0 \subseteq D(S)$ and $S_0$ is closed under star, so that $S_0$ is a Conway subsemiring of $S$. We know that any $s \in \mathbf{Rat}_S(S_0, \Sigma)$ can be written as a sum $x + a$, where $x \in S_0$ and $a \in \mathbf{Rat}'_S(S_0, \Sigma) \subseteq D(S)$. Now $S_0 \subseteq D(S)$ by assumption, and since $D(S)$ is an ideal containing both $x$ and $a$, it follows that $s = x + a \in D(S)$. Since $S$ is a partial Conway semiring, $s^* = (x^*a)^*x^* = x^* + (x^*a)^+x^*$. By assumption, $x^* \in S_0$. Also, $x^*a \in \mathbf{Rat}'_S(S_0, \Sigma)$, since $\mathbf{Rat}'_S(S_0, \Sigma)$ is closed under multiplication with elements of $S_0$. Thus, since $\mathbf{Rat}'_S(S_0, \Sigma)$ is closed under plus and multiplication with elements of $S_0$, we have that $(x^*a)^+x^* \in \mathbf{Rat}'_S(S_0, \Sigma)$. Since $s^*$ is the sum of an element of $S_0$ and an element of $\mathbf{Rat}'_S(S_0, \Sigma)$, it follows that that $s^*$ is in $\mathbf{Rat}_S(S_0, \Sigma)$.

Note that when $S_0 \subseteq D(S)$ and $S_0$ is closed under star, then, by the above argument, $\mathbf{Rat}_S(S_0, \Sigma) \subseteq D(S)$ is also closed under star, so that it is a Conway semiring.

Next, assume that (23) holds. Let $s = x + a \in \mathbf{Rat}_S(S_0, \Sigma)$, where $x \in S_0$ and $a \in \mathbf{Rat}'_S(S_0, \Sigma)$. We want to show that if $s$ is in $D(S)$, then $s^+$ is also in $\mathbf{Rat}_S(S_0, \Sigma)$. But by (23), $s \in D(S)$ only if $x = 0$. In that case, $s^+ = a^+ \in \mathbf{Rat}'_S(S_0, \Sigma) \subseteq \mathbf{Rat}_S(S_0, \Sigma)$. □

REMARK 6.7 *Note that the second condition in the above proposition holds whenever each $s \in S$ has at most one representation $s = x + a$ with $x \in S_0$ and $a \in D(S)$. This happens when $S$ is the direct sum of $S_0$ and $D(S)$.*

In the proof of our Kleene theorem, we will make use of the following fact.

LEMMA 6.8 *Suppose that each entry of the $n \times n$ matrix $A$ is in $\mathbf{Rat}'_S(S_0, \Sigma)$. Then the same holds for the matrix $A^+$.*

*Proof.* We prove this fact by induction on $n$. When $n = 0$ or $n = 1$, our claim is clear. Assuming that $n > 0$ write $A = \begin{pmatrix} a & b \\ c & d \end{pmatrix}$, where $a$ is $(n-1) \times (n-1)$, $d$ is $1 \times 1$. Then $A^+$ is given by the formula (17). We only show that each entry of the submatrix $(a + bc^*d)^+$ is in $\mathbf{Rat}'_S(S_0, \Sigma)$. But $a + bc^*d = a + bd + bc^+d$. By the induction hypothesis, each entry of $c^+$ is in $\mathbf{Rat}'_S(S_0, \Sigma)$. Since $\mathbf{Rat}'_S(S_0, \Sigma)$ is closed under sum and product, and since each entry of $a, b$ or $d$ is also in this set, it follows that each entry of $a + bc^*d$ is in $\mathbf{Rat}'_S(S_0, \Sigma)$. Thus, using the induction hypothesis again, it follows that each entry of $(a + bc^*d)^+$ is in $\mathbf{Rat}'_S(S_0, \Sigma)$. □

PROPOSITION 6.9 *For any $S, S_0$ and $\Sigma$ as above, $\mathbf{Rec}_S(S_0, \Sigma) \subseteq \mathbf{Rat}_S(S_0, \Sigma)$.*

*Proof.* Let $\mathbf{A} = (\alpha, A, \beta)$ be an automaton over $(S_0, \Sigma)$. Then $|\mathbf{A}| = \alpha A^* \beta = \alpha \beta + \alpha A^+ \beta$. Clearly, $\alpha \beta \in S_0$. By the previous lemma, $A^+ \in \mathbf{Rat}'_S(S_0, \Sigma)$. Since $\mathbf{Rat}'_S(S_0, \Sigma)$ is closed under left and right multiplication with elements of $S_0$ and since $\mathbf{Rat}'_S(S_0, \Sigma)$ is closed under sum, it follows that $\alpha A^+ \beta$ is in $\mathbf{Rat}'_S(S_0, \Sigma)$. Thus, $|\mathbf{A}|$ is the sum of an element of $S_0$ and an element of $\mathbf{Rat}'_S(S_0, \Sigma)$, showing that $|\mathbf{A}|$ is in $\mathbf{Rat}_S(S_0, \Sigma)$. □

We now prove the converse of the previous proposition.

PROPOSITION 6.10 *For any $S, S_0$ and $\Sigma$ as above, $\mathbf{Rat}_S(S_0, \Sigma) \subseteq \mathbf{Rec}_S(S_0, \Sigma)$.*

*Proof.* Suppose that $s \in \mathbf{Rat}_S(S_0, \Sigma)$. We have to show that there is an automaton over $(S_0, \Sigma)$ whose behavior is $s$. First we prove this claim for the elements of $\mathbf{Rat}'_S(S_0, \Sigma)$. We show: For each $s \in \mathbf{Rat}'_S(S_0, \Sigma)$ there exists an automaton over $(S_0, \Sigma)$ whose behavior is $s$ such that *the product of the initial and the final vector of $\mathbf{A}$ is* 0. Assume that $s = 0$. Then consider the automaton $\mathbf{A}_0 = (0, 0, 0)$ of dimension 1. We have that $|\mathbf{A}_0| = 0$ and it is clear that the product of the initial and the final vector is 0. Next let $s = a$ for some letter $a \in \Sigma$. Then define the following automaton $\mathbf{A}_a$ of dimension 2:

$$\mathbf{A}_a = \left( \begin{pmatrix} 1 & 0 \end{pmatrix}, \begin{pmatrix} 0 & a \\ 0 & 0 \end{pmatrix}, \begin{pmatrix} 0 \\ 1 \end{pmatrix} \right).$$

We have

$$|\mathbf{A}_a| = \begin{pmatrix} 1 & 0 \end{pmatrix} \begin{pmatrix} 1 & a \\ 0 & 1 \end{pmatrix} \begin{pmatrix} 0 \\ 1 \end{pmatrix} = a.$$

In the induction step there are five cases to consider. Suppose that $s = s_1 + s_2$ or $s = s_1 s_2$ such that there exist automata $\mathbf{A}_i = (\alpha_i, A_i, \beta_i)$ over $(S_0, \Sigma)$ with $|\mathbf{A}_i| = s_i$ satisfying $\alpha_i \beta_i = 0$, $i = 1, 2$. We construct automata $\mathbf{A}_1 + \mathbf{A}_2$, $\mathbf{A}_1 \cdot \mathbf{A}_2$ defining $s_1 + s_2$ and $s_1 s_2$, respectively. Let

$$\mathbf{A}_1 + \mathbf{A}_2 = \left( (\alpha_1, \alpha_2), \begin{pmatrix} A_1 & 0 \\ 0 & A_2 \end{pmatrix}, \begin{pmatrix} \beta_1 \\ \beta_2 \end{pmatrix} \right)$$

and

$$\mathbf{A}_1 \cdot \mathbf{A}_2 = \left( (\alpha_1, 0), \begin{pmatrix} A_1 & \beta_1 \alpha_2 A_2 \\ 0 & A_2 \end{pmatrix}, \begin{pmatrix} \beta_1 \alpha_2 \beta_2 \\ \beta_2 \end{pmatrix} \right).$$

Then

$$\begin{aligned} |\mathbf{A}_1 + \mathbf{A}_2| &= (\alpha_1, \alpha_2) \begin{pmatrix} A_1^* & 0 \\ 0 & A_2^* \end{pmatrix} \begin{pmatrix} \beta_1 \\ \beta_2 \end{pmatrix} \\ &= \alpha_1 A_1^* \beta_1 + \alpha_2 A_2^* \beta_2 \\ &= |\mathbf{A}_1| + |\mathbf{A}_2|, \end{aligned}$$

and

$$\begin{aligned} |\mathbf{A}_1 \cdot \mathbf{A}_2| &= (\alpha_1, 0) \begin{pmatrix} A_1^* & A_1^* \beta_1 \alpha_2 A_2^+ \\ 0 & A_2^* \end{pmatrix} \begin{pmatrix} \beta_1 \alpha_2 \beta_2 \\ \beta_2 \end{pmatrix} \\ &= \alpha_1 A_1^* \beta_1 \alpha_2 \beta_2 + \alpha_1 A_1^* \beta_1 \alpha_2 A_2^+ \beta_2 \\ &= \alpha_1 A_1^* \beta_1 \alpha_2 A_2^* \beta_2 \\ &= |\mathbf{A}_1| \cdot |\mathbf{A}_2|. \end{aligned}$$

Also,

$$(\alpha_1, \alpha_2)(\beta_1, \beta_2)^T = \alpha_1 \beta_1 + \alpha_2 \beta_2 = 0$$

and
$$(\alpha_1, 0)(\beta_1\alpha_2\beta_2, \beta_2)^T = \alpha_1\beta_1\alpha_2\beta_2 = 0.$$

(Of course, we could have used the fact that $\alpha_2\beta_2 = 0$ earlier in the definition of $\mathbf{A}_1 \cdot \mathbf{A}_2$, but we wanted to show that the construction works even if this does not hold.)

Next, we show that when $s = r^+$ for some $r$ which is the behavior of an automaton $\mathbf{A} = (\alpha, A, \beta)$ over $(S_0, \Sigma)$, such that $\alpha\beta = 0$, then $s$ is the behavior of an automaton $\mathbf{A}^+$. Since

$$|\mathbf{A}| = \alpha A^* \beta = \alpha\beta + \alpha A^+ \beta = \alpha A^+ \beta$$

thus $r = \alpha A^* \beta = \alpha A^+ \beta$. Now let

$$\mathbf{A}^+ = (\alpha, A + \beta\alpha A, \beta).$$

By $(A + \beta\alpha A)^* = A^*(\beta\alpha A^+)^*$, we have

$$|\mathbf{A}^+| = \alpha A^*(\beta\alpha A^+)^*\beta = \alpha A^+ \beta(\alpha A^+ \beta)^* = (\alpha A^+ \beta)^+ = |\mathbf{A}|^+ = s.$$

By assumption, we have that $\alpha\beta = 0$.

Last, if $s = |\mathbf{A}|$ and $x \in S_0$, where $\mathbf{A} = (\alpha, A, \beta)$ is an automaton over $(S_0, \Sigma)$ with $\alpha\beta = 0$, then $xs = |x\mathbf{A}|$ and $sx = |\mathbf{A}x|$ where $x\mathbf{A} = (x\alpha, A, \beta)$ and $\mathbf{A}x = (\alpha, A, \beta x)$. Also $(x\alpha)\beta = \alpha(\beta x) = 0$.

We have thus shown that $\mathbf{Rat}'_S(S_0, \Sigma) \subseteq \mathbf{Rec}_S(S_0, \Sigma)$. Finally, if $s \in \mathbf{Rat}_S(S_0, \Sigma)$, so that $s = x + a$ for some $x \in S_0$ and $a \in \mathbf{Rat}'_S(S_0, \Sigma)$, then there is an automaton $\mathbf{A} = (\alpha, A, \beta)$ over $(S_0, \Sigma)$ whose behavior is $a$. Then define $\mathbf{B} = \left( \begin{pmatrix} x & \alpha \end{pmatrix}, \begin{pmatrix} 0 & 0 \\ 0 & A \end{pmatrix}, \begin{pmatrix} 1 \\ \beta \end{pmatrix} \right)$. It holds that

$$|\mathbf{B}| = \begin{pmatrix} x & \alpha \end{pmatrix} \begin{pmatrix} 1 & 0 \\ 0 & A^* \end{pmatrix} \begin{pmatrix} 1 \\ \beta \end{pmatrix} = x + \alpha A^* \beta = x + a = s.$$

$\square$

We have proved:

THEOREM 6.11 *Suppose that $S$ is a partial Conway semiring, $S_0$ is a subsemiring of $S$ and $\Sigma \subseteq D(S)$. Then $\mathbf{Rec}_S(S_0, \Sigma) = \mathbf{Rat}_S(S_0, \Sigma)$.*

COROLLARY 6.12 *Suppose that $S$ is a Conway semiring, $S_0$ is a Conway subsemiring of $S$ and $\Sigma \subseteq S$. Then $\mathbf{Rec}_S(S_0, \Sigma) = \mathbf{Rat}_S(S_0, \Sigma)$ is the least Conway subsemiring of $S$ which contains $S_0 \cup \Sigma$.*

COROLLARY 6.13 *Suppose that $S$ is a partial Conway semiring, $S_0$ is a subsemiring of $S$ and $\Sigma \subseteq D(S)$. Suppose that whenever $x + a \in D(S)$ for some $x \in S_0$ and $a \in D(S)$ then $x = 0$. Then $\mathbf{Rec}_S(S_0, \Sigma) = \mathbf{Rat}_S(S_0, \Sigma)$ is the least partial Conway subsemiring of $S$ which contains $S_0 \cup \Sigma$.*

The case when the partial Conway semiring is a power series semiring deserves special attention. Let $S$ be a semiring and $\Sigma$ a set, and consider the partial iteration semiring $S\langle\!\langle\Sigma^*\rangle\!\rangle$. Recall that the star operation is defined on the proper power series and that $S$ can be identified with a subsemiring of $S\langle\!\langle\Sigma^*\rangle\!\rangle$. We denote $\mathbf{Rat}_{S\langle\!\langle\Sigma^*\rangle\!\rangle}(S, \Sigma)$ by $S^{\mathrm{rat}}\langle\!\langle\Sigma^*\rangle\!\rangle$ and $\mathbf{Rec}_{S\langle\!\langle\Sigma^*\rangle\!\rangle}(S, \Sigma)$ by $S^{\mathrm{rec}}\langle\!\langle\Sigma^*\rangle\!\rangle$. Since $S\langle\!\langle\Sigma^*\rangle\!\rangle$ is the direct sum of $S$ and the ideal of proper power series, (23) is satisfied. Thus, $S^{\mathrm{rat}}\langle\!\langle\Sigma^*\rangle\!\rangle$ is closed under the star operation, and thus $S^{\mathrm{rat}}\langle\!\langle\Sigma^*\rangle\!\rangle$ is partial iteration semiring.

COROLLARY 6.14 *Suppose that $S$ is a semiring and $\Sigma$ a set. Then $S^{\mathrm{rat}}\langle\!\langle \Sigma^*\rangle\!\rangle$ is the least partial iteration subsemiring of $S\langle\!\langle \Sigma^*\rangle\!\rangle$ containing $S \cup \Sigma$. Moreover, $S^{\mathrm{rat}}\langle\!\langle \Sigma^*\rangle\!\rangle = S^{\mathrm{rec}}\langle\!\langle \Sigma^*\rangle\!\rangle$.*

Recall from Remark 5.11 that when $S$ is a Conway or iteration semiring, then so is $S\langle\!\langle \Sigma^*\rangle\!\rangle$, for any set $\Sigma$.

COROLLARY 6.15 *Suppose that $S$ is a Conway semiring. Then $S^{\mathrm{rat}}\langle\!\langle \Sigma^*\rangle\!\rangle$ is the least Conway subsemiring of $S\langle\!\langle \Sigma^*\rangle\!\rangle$ containing $S \cup \Sigma$. Moreover, $S^{\mathrm{rat}}\langle\!\langle \Sigma^*\rangle\!\rangle = S^{\mathrm{rec}}\langle\!\langle \Sigma^*\rangle\!\rangle$.*

The following result is used in [4].

THEOREM 6.16 *Suppose that $S$ is a semiring, $\Sigma$ is a set, and $S'$ is a partial Conway semiring. Then a function $h : S^{\mathrm{rat}}\langle\!\langle \Sigma^*\rangle\!\rangle \to S'$ is a morphism of partial Conway semirings iff the following hold.*

1. *The restriction of $h$ onto $S$ is a semiring morphism.*
2. $\Sigma h \subseteq D(S')$.
3. *$h$ preserves linear combinations in $S\langle \Sigma\rangle$, i.e., $(s_1 a_1 + \cdots + s_n a_n)h = (s_1 h)(a_1 h) + \cdots + (s_n h)(a_n h)$ for all $s_i \in S$, $a_i \in \Sigma$, $i = 1, \ldots, n$, $n \geq 0$.*
4. *$h$ preserves the behavior of automata: For every automaton $\mathbf{A} = (\alpha, A, \beta)$ in $S^{\mathrm{rat}}\langle\!\langle \Sigma^*\rangle\!\rangle$, $|\mathbf{A}|h = |\mathbf{A}h|$, where $\mathbf{A}h$ is the automaton $(\alpha h, Ah, \beta h)$ over $(Sh, \Sigma h)$ in $S'$.*

*Proof.* It is clear that the conditions are necessary. Suppose now that $h$ satisfies the above conditions. Since the restriction of $h$ onto $S$ is a semiring morphism, $h$ preserves the constants 0 and 1. To prove that $h$ preserves sum, consider rational series $s_1 = |\mathbf{A}_1|$ and $s_2 = |\mathbf{A}_2|$ in $S^{\mathrm{rat}}\langle\!\langle \Sigma^*\rangle\!\rangle$, where $\mathbf{A}_i = (\alpha_i, A_i, \beta_i)$ are automata over $(S_0, \Sigma)$ for $i = 1, 2$. Let $\mathbf{A} = \mathbf{A}_1 + \mathbf{A}_2$ be defined as in the proof of Theorem 6.11. Since $h$ maps $\Sigma$ into $D(S')$ and preserves linear combinations in $S\langle\Sigma\rangle$, we have that $\mathbf{A}h = (\alpha h, Ah, \beta h)$ is an automaton over $(S_0 h, \Sigma h)$. Since $h$ preserves behavior of automata,

$$\begin{aligned}(s_1 + s_2)h &= |\mathbf{A}|h \\ &= |\mathbf{A}h| \\ &= |(\mathbf{A}_1 + \mathbf{A}_2)h| \\ &= |\mathbf{A}_1 h + \mathbf{A}_2 h| \\ &= |\mathbf{A}_1 h| + |\mathbf{A}_2 h| \\ &= |\mathbf{A}_1|h + |\mathbf{A}_2|h \\ &= s_1 h + s_2 h.\end{aligned}$$

The fact that $(s_1 s_2)h = (s_1 h)(s_2 h)$ can be proved in the same way using the construction of the automaton $\mathbf{A}_1 \cdot \mathbf{A}_2$. Last, we prove that $h$ preserves $^+$. For this reason, let $s$ be a proper rational series in $S^{\mathrm{rat}}\langle\!\langle \Sigma^*\rangle\!\rangle$. Let $\mathbf{A} = (\alpha, A, \beta)$ be an automaton over $(S, \Sigma)$ whose behavior $\alpha A^* \beta = \alpha\beta + \alpha A^+ \beta = \alpha A^+ \beta$ is $s$. Consider the automaton $\mathbf{A}^+$ defined in the proof of Theorem 6.11. Then, $|\mathbf{A}^+| = |\mathbf{A}|^+$ and $|(\mathbf{A}h)^+| = |\mathbf{A}h|^+$. Thus, since $h$ preserves behavior,

$$\begin{aligned}s^+ h &= |\mathbf{A}^+|h \\ &= |\mathbf{A}^+ h| \\ &= |(\mathbf{A}h)^+| \\ &= |\mathbf{A}h|^+ \\ &= (sh)^+.\end{aligned}$$

# References

[1] J. Berstel and Ch. Reutenauer, *Rational Series and Their Languages*, Springer, 1988. Verison of October 19, 2007, http://www-igm.univ-mlv.fr/∼berstel/.

[2] S.L. Bloom and Z. Ésik, Matrix and matricial iteration theories, Part I, *J. Comput. Sys. Sci.*, 46(1993), 381–408.

[3] S.L. Bloom and Z. Ésik, *Iteration Theories: The Equational Logic of Iterative Processes*, EATCS Monographs on Theoretical Computer Science, Springer–Verlag, 1993.

[4] S.L. Bloom and Z. Ésik, Axiomatizing rational power series, to appear.

[5] S. Bloom, S. Ginali and J.D. Rutlege, Scalar and vector iteration, *J. Computer System Sciences*, 14(1977), 251–256.

[6] J.C. Conway. *Regular Algebra and Finite Machines*, Chapman and Hall, London, 1971.

[7] M. Droste and W. Kuich, Semirings and formal power series, in: *Handbook of Weighted Automata*, Springer, to appear.

[8] S. Eilenberg, *Automata, Languages, and Machines*. vol. A, Academic Press, 1974.

[9] C.C. Elgot, Monadic computation and iterative algebraic theories, in: *Logic Colloquium 1973, Studies in Logic*, J.C. Shepherdson, editor, volume 80, North Holland, Amsterdam, 1975, 175–230.

[10] C.C. Elgot, Matricial theories, *J. of Algebra*, 42(1976), 391–421.

[11] Z. Ésik, Group axioms for iteration, *Information and Computation*, 148(1999), 131–180.

[12] Z. Ésik and W. Kuich, Inductive $^*$-semirings, *Theoret. Comput. Sci.*, 324(2004), 3–33.

[13] Z. Ésik and W. Kuich, Equational axioms for a theory of automata, in: *Formal Languages and Applications*, Studies in Fuzziness and Soft Computing 148, Springer, 2004, 183–196.

[14] J.S. Golan, *The Theory of Semirings with Applications in Computer Science*, Longman Scientific and Technical, 1993.

[15] G. Grätzer, *Universal Algebra*, Springer, 1979.

[16] D. Krob, Complete systems of B-rational identities, *Theoretical Computer Science*, 89(1991), 207–343.

[17] D. Krob, Matrix versions of aperiodic $K$-rational identities, *Theoretical Informatics and Appllications*, 25(1991), 423–444.

[18] V.N. Redko, On the determining totality of relations of an algebra of regular events (in Russian), *Ukrainian Math. Ž.*, 16(1964), 120–126.

[19] V.N. Redko, On algebra of commutative events (in Russian), *Ukrainian Math. Ž.*, 16(1964), 185–195.